\documentclass[12pt]{article}
\usepackage{epsfig,amssymb}
\usepackage{latexsym}

\newcommand{\bq}{\begin{equation}}
\newcommand{\eq}{\end{equation}}
\newcommand{\bqa}{\begin{eqnarray}}
\newcommand{\eqa}{\end{eqnarray}}
\newcommand{\ben}{\begin{enumerate}}
\newcommand{\een}{\end{enumerate}}
\newcommand{\bc}{\begin{center}}
\newcommand{\ec}{\end{center}}
\newcommand{\bqb}{\begin{eqnarray*}}
\newcommand{\eqb}{\end{eqnarray*}}

\def\pr#1#2#3{ Phys. Rev. ${\bf{#1}}$:#2 (#3)}
\def\prl#1#2#3{ Phys. Rev. Lett. ${\bf{#1}}$:#2 (#3)}
\def\pl#1#2#3{ Phys. Lett. ${\bf{#1}}$:#2 (#3)}

\def\np#1#2#3{ Nucl. Phys. ${\bf{#1}}$:#2 (#3)}
\def\zp#1#2#3{ Z. f. Phys. ${\bf{#1}}$:#2 (#3)}

\def\epj#1#2#3{ Eur. Phys. J. ${\bf{#1}}$:#2 (#3)}
\def\ijmp#1#2#3{ Int. J. Mod. Phys. ${\bf{#1}}$:#2 (#3)}

\def\eg{{\it e.g. }}

\def\etal{{\it et.al. }}

\def\L{ {\cal L }}
\def\Op{ {\cal O}}
\def\sw{s_W}
\def\cw{c_W}
\def\swd{s^2_W}
\def\cwd{c^2_W}
\def\ed{e^2}
\def\mwd{m_W^2}
\def\mw{m_W}

\def\mzd{m_Z^2}

\def\tchi{\tilde \chi}

\def\tt{\tau_3}
\def\costheta{\cos\theta}
\def\sintheta{\sin\theta}
\def\lw{\alpha_W}
\def\lsm{\biggm( {\lw s\over\mw}\biggm) }
\def\rd{\sqrt2}

\global\nulldelimiterspace = 0pt

\begin{document}
\pagenumbering{arabic}
\thispagestyle{empty}
\def\thefootnote{\fnsymbol{footnote}}
\setcounter{footnote}{1}

\begin{flushright}
September 2005\\
hep-ph/0510061\\

 \end{flushright}

\vspace{2cm}
\begin{center}
{\Large\bf Contrasting the anomalous and the SM-MSSM couplings
at the Colliders\footnote{Partially supported by the
Greek Ministry of Education and Religion and the
EPEAK program Pythagoras}}.  \vspace{1.cm}  \\
{\large G.J. Gounaris }\\
\vspace{0.2cm}
Department of Theoretical Physics, Aristotle
University of Thessaloniki,\\
Gr-54124, Thessaloniki, Greece.\\

\vspace*{1.cm}

{\bf Abstract}
\end{center}
\vspace*{-0.4cm}
This talk consists of two parts. In the first,
 the present experimental
bounds on the anomalous couplings of the gauge bosons,
based mainly on the LEP and Tevatron experiments,  are reviewed.
In the second part, the theorem of
helicity conservation (HC) is presented, which should
be valid in  either the Standard Model (SM) or MSSM, for
 any two-body process at high energies and fixed angles.
 The energy-range for the HC validity
  is discussed and, under certain conditions, it
   should well be within the  LHC or ILC range.
  Since all known anomalous couplings  violate HC,
  its testing may  provide a way for generically identifying the
  possible presence of anomalous (non-renormalizable) contributions.

\vspace*{1.cm}

\begin{center}
PRESENTED\\
\vspace*{0.2cm}
at the 2005 Photon Linear Collider Workshop (PLC2005)\\
Kazimierz, Poland, 5-8 September, 2005.

\end{center}
\vspace*{-0.4cm}

\vspace{0.5cm}

\def\thefootnote{\arabic{footnote}}
\setcounter{footnote}{0}
\clearpage

\section{Introduction}

The description of particle physics through renormalizable
$SU(3)\times SU(2)\times U(1)$
gauge invariant interactions, has been  impressively successful,
up to now.

The keyword here is
\underline{renormalizable}, which imposes that only operators
of dimensions less than or equal to four, can appear in the Lagrangian.
This property, together with the group structure, determine
the gauge and matter interactions,  leading \eg to the
most striking phenomenon of {\it asymptotic freedom} which
permeates contemporary particle and cosmology physics  \cite{AsFree}.

In order to thoroughly test experimentally  these interactions,
alternative models are   envisaged, which may be used as
 parameterizations of any possible violation of their validity.
As such, in the present context we consider anomalous gauge
couplings \cite{GG, Hagiwara, TGC}.
These anomalous couplings can always be assumed
to obey $SU(3)\times SU(2)\times U(1)$
symmetry \cite{GR}; but, due to their higher dimensionality,
they violate {\it renormalizability}.

Extensive phenomenological
studies have already been made in  various  specific processes,
comparing  the signatures of  such
couplings, to those of \eg the Standard Model (SM). On the basis of these,
 experimental searches have  been performed at LEP,
the Tevatron and elsewhere; which invariably  impose ever
growing constraints on the magnitude of any conceivable anomalous
coupling. Thus, at present at least,
SM  (as well as  its renormalizable SUSY extensions), are  fully consistent
with Nature.

The strength of these constraints will most probably further
increase when LHC or  ILC
 start operating, basically because the non-renormalizable nature
of the anomalous couplings bounds  their  effects to increase
strongly with energy. Such a strong increase
 is in fact  a common feature
 of  all effectively non-renormalizable ways of going beyond SM
 or its SUSY  analogs\footnote{Similar    effects are
   observed  \eg in   extra large dimension models determined by an effectively
    non-renormalizable lagrangian. }.
 In turn, this  facilitates
 their exclusion, provided of course we adhere to the usual practice
 of considering \eg only a few anomalous couplings at a time.

 As the energy increases reaching the LHC range though,
it becomes increasingly difficult  to motivate the idea that
the anomalous couplings may   be parameterized  by a
few {\it dimension=6} operators only.
Instead, higher dimensional operators (as well as previously ignored
{\it dim=6} ones) should  be considered together;
particularly if the scale of new physics is reached there,
thereby  seriously  reducing the
ability to  constrain the anomalous  couplings.\\

A partial  solution to this difficulty is offered
by the    property
called helicity conservation (HC),
which in SM and its renormalizable SUSY extensions,
greatly reduces the number of
non-vanishing amplitudes at very high energies and fixed angles \cite{helicity}.
Combining this with the observation that
   \underline{all known} anomalous couplings
violate HC, we obtain  a generic test for all of them.

The importance of HC as a property of  SM and MSSM,
and in fact of any renormalizable gauge theory, can hardly be overemphasized.
Its validity, particularly for gauge amplitudes  in SM,
is only established after large cancellation from
 different  diagrams, which are only realized for renormalizable
 couplings \cite{helicity}.
 Because of this, HC  is   not directly obvious from the SM Lagrangian,
 and it must  somehow be related to the twistor structure in QCD \cite{twistor}.
The possible appearance of  HC violation  indicates the presence of some
non-renormalizable contributions, an  example of which is
of course the anomalous couplings \cite{helicity}.

In the first part of this talk I review the present
constraints on the anomalous gauge couplings; while in the second part,
HC is described.\\

\section{Anomalous electroweak   couplings}

As is well known, anomalous electroweak couplings may be introduced in SM
or MSSM by including operators of higher than four dimension,
which  preserve the $SU(3)\times SU(2)\times U(1)$ gauge symmetry.
These operators  induce anomalous couplings  not only to the gauge
bosons, but also to the   Higgs particles \cite{Higgs}, and the quarks
and leptons, particularly of the third family \cite{top}.
Since no Higgs particle has yet been discovered,
and the top anomalies  are covered by J. Wudka \cite{Wudka}, we will concentrate
here on the purely gauge anomalous couplings.\\

\subsection{$W^\pm$ anomalous Couplings}
The most general set  of the  anomalous triple gauge couplings (TGC)
describing all possible  $(W^+W^-Z)$ and $(W^+W^-\gamma)$ vertices,
is  traditionally parameterized as \cite{GG, Hagiwara, TGC}
\bqa
\L^{\rm TGC}_{NP}& =&-i e g_{VWW} \Big \{
(1+\delta g_1^V) V^\mu (W^-_{\mu\nu}W^{+\nu}-W^+_{\mu\nu}W^{-\nu})
+(1+\delta \kappa_V)V^{\mu\nu} W^+_\mu W^-_\nu \nonumber \\
&+&\frac{\lambda_V}{\mwd}V^{\mu\nu} W^{+\rho}_\nu W^-_{\rho\mu}
+ig_5^V\epsilon_{\mu\nu\rho\sigma}
[(\partial^\rho W^{-\mu})W^{+\nu}- W^{-\mu}(\partial^\rho W^{+\nu})]V^\sigma
\nonumber \\
&+& i g_4^V W^-_\mu W^+_\nu (\partial^\mu V^\nu+\partial^\nu V^\mu)
-\tilde \kappa_V W^-_\mu W^+_\nu \tilde V_{\mu\nu}-
\frac{\tilde \lambda_V}{\mwd} W^-_{\rho\mu}W^{+\mu}\tilde V^{\nu\rho} \Big \}
~, \label{W-TGC}
\eqa
where
\bqa
&& \tilde V_{\mu\nu}=\frac{1}{2}\epsilon_{\mu\nu\rho\sigma} V^{\rho\sigma}~~,~~
\nonumber \\
&&  V=\gamma ~,~ Z  ~~~~ \leftrightarrow ~~~~
 g_{\gamma WW}=1~,~ g_{ZWW}=\frac{\cw}{\sw}~~. \label{W-TGC1}
\eqa
The anomalous couplings $(\delta g_1^V~,~\delta\kappa_V~,~\lambda_V~,~ g_5^V)$
respect CP, while $(g_4^V~,~\tilde \kappa_V~,~\tilde \lambda_V)$ violate it.
For the photon couplings in particular, $U_{\rm em}(1)$ gauge invariance
implies that
\[
\delta g_1^\gamma \sim \frac{q^2}{\Lambda^2}~~,~~
g_5^\gamma \sim \frac{q^2}{\Lambda^2}~~,
~~g_4^\gamma \sim \frac{q^2}{\Lambda^2}~~,
\]
as the off-shell photon approaches its mass shell value $q^2=0$.

The phases in the effective lagrangian (\ref{W-TGC}) have been chosen
so that all couplings are real, in case the scale of
the new physics (NP)  inducing them is very high.
If the NP scale is low though,
pole and branch-point singularities  develop.

All anomalous TGC are consistent with
 $SU(3)\times SU(2)\times U(1)$ gauge invariance, provided they are combined
 with appropriate interactions involving more gauge and/or physical Higgs particles.
 To achieve this for the actual couplings in (\ref{W-TGC}) though,
 operators of dimension up to 12 need be considered \cite{GR}.

 Of course, if the NP scale is not very high, like
 \eg in a  SUSY case with the new particles at the LHC range,
  operators of any  dimension  would be  allowed, seriously weakening
 our ability to constrain them.

 If, on the contrary, the NP scale is
 high though, and the physical Higgs particles are within the
 electroweak  range, then the natural couplings of the induced operators should be
$ g_0  \sim 1/ \Lambda^{{\rm dim}-4}$,
allowing  the contemplation that dimension=6
operators\footnote{Alternative ways of  ordering  the NP operators
have been contemplated, in case no light Higgs particles exist;
see \eg \cite{Yuan}. }  could  be sufficient in describing NP.

Disregarding  all such operators which are strongly excluded due
to their tree-level contributions to physical observables, and assuming also
that only one SM-like light Higgs particle exists,  we  parameterize the
anomalous contribution to the effective lagrangian describing
the $W^\pm$ TGC as \cite{Stong}
\bqa
\L^{\rm TGC}_{NP}({\rm dim} =6)& =& \frac{e}{\cw \mwd} \alpha_{B\phi}\Op_{B\phi}
+\frac{e}{\sw \mwd} \alpha_{W\phi}\Op_{W\phi}+
\frac{e}{\sw \mwd} \alpha_{W}\Op_{W} \nonumber \\
&+& \frac{e^2}{2 \sw \cw \mwd}\tilde \alpha_{BW}\tilde \Op_{BW}
+ \frac{e}{ \sw  \mwd}\tilde \alpha_{W}\tilde \Op_{W} ~~, \label{W-TGC-d6}
\eqa
with
\bqa
\Op_W=\frac{1}{3!}  (\vec W_{\mu\nu} \times \vec W^{\nu\lambda})
\cdot \vec W_\lambda^{~\mu} ~~, &
\Op_{W\phi}=i D_\mu \phi^\dagger \vec \tau \vec W^{\mu\nu} D_\nu\phi & ~,~~
\Op_{B\phi}=i D_\mu \phi^\dagger  B^{\mu\nu} D_\nu\phi ~~, \nonumber \\
\tilde \Op_W=\frac{1}{3!}  (\vec W_{\mu\nu} \times \vec W^{\nu\lambda})
\cdot \tilde {\vec W}_\lambda ^{~\mu} ~~, &
\tilde \Op_{W\phi}=\frac{i}{2}  \phi^\dagger \vec \tau \tilde{ \vec W}^{\mu\nu}
\phi  B_{\mu\nu} & ~, \label{W-TGC-op}
\eqa
where the first three operators conserve CP, while the rest two violate it.
The anomalous couplings defined in (\ref{W-TGC}), are related to those
in (\ref{W-TGC-d6}), by
\bqa
&&  \delta g_1^Z=\frac{\alpha_{W\phi}}{\cwd} ~~,~~
\lambda_\gamma=\lambda_Z=\alpha_W ~~,~~
\delta \kappa_\gamma =-\frac{\cwd}{\swd}(\delta\kappa_Z-\delta g_1^Z)=
\alpha_{W\phi}+\alpha_{B\phi}~~, \nonumber \\
&& \tilde \kappa_\gamma=-\frac{\cwd}{\swd}\tilde \kappa_Z=\tilde \alpha_{BW}~~,~~
\tilde \lambda_\gamma=\tilde \lambda_Z=\tilde \alpha_W ~~.\label{d6-cons1}
\eqa

\begin{figure}[htb]
\vspace*{-1.5cm}
\[
\hspace{-0.5cm}\epsfig{file=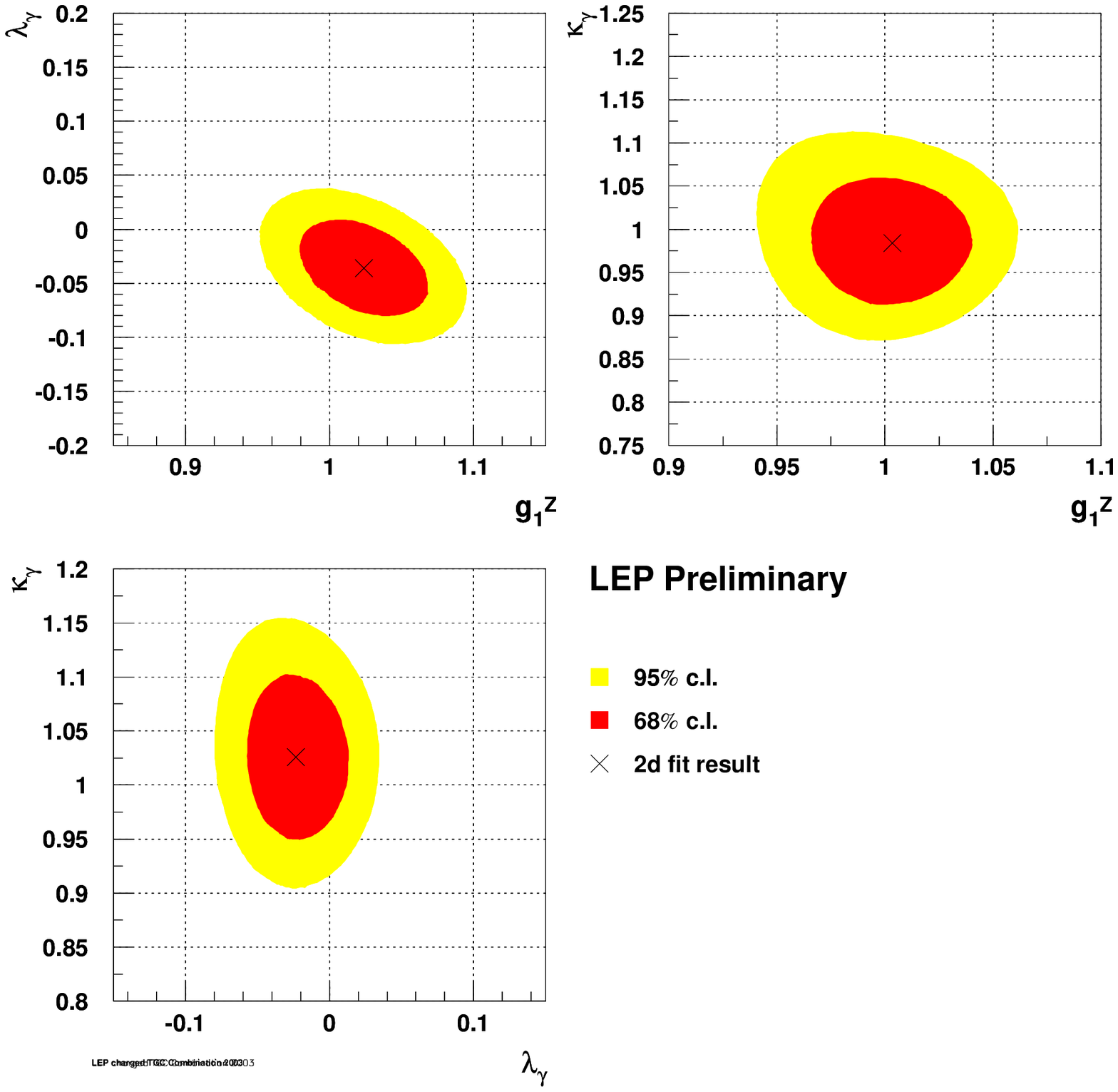,height=13.cm, width=13cm}
\]
\vspace*{-1.5cm}
\caption[1]{The  combined LEP2 results from \cite{LEPres}. In each
case two of the parameters in (\ref{d6-cons2}) are varied,
while the third is fixed at its standard value.}
\label{LEP-W-TGCfig}
\end{figure}

Restricting to CP conserving couplings only, and using the definitions
\[
\kappa_\gamma\equiv 1+\delta \kappa_\gamma~~,~~
\kappa_Z \equiv 1+\delta \kappa_Z  ~~,~~
g_1^V \equiv 1+\delta g_1^V ~~,~~
\]
 we end up in a situation where
only the  three independent couplings
\bq
~~ g_1^Z ~~,~~ \kappa_\gamma ~~,~~ \lambda_\gamma \label{d6-cons2}
\eq
participate, whose standard  values are (1,1,0) respectively. The
fitted LEP ranges for these parameters from \cite{LEPres} are indicated in
Fig.\ref{LEP-W-TGCfig} and Table 1, obtained respectively by
varying two or one parameter at a time.
\begin{table}[htbp]
\begin{center}
\begin{tabular}{|l||r|c|}
\hline
Parameter  & 68\% C.L.   & 95\% C.L.      \\
\hline
\hline
$g_1^Z$     & $0.991^{0.022}_{0.021} $  & [$0.949,~~1.034$]  \\
\hline
$\kappa_\gamma$     & $0.984^{0.042}_{0.047}$  & [$0.895,~~1.069$]  \\
\hline
$\lambda_\gamma$     & $-0.016^{0.021}_{0.023}$  & [$-0.059,~~0.026$]  \\
\hline
\end{tabular}
\caption[]{ The  combined LEP2 results from \cite{LEPres}.  In
  each case the listed parameter  is varied while the other two
  of (\ref{d6-cons2}) are   fixed to their standard  values.}
\end{center}
\label{LEP-W-TGCtab}
\end{table}

The corresponding one-parameter
Tevatron D0 fits  from \cite{D0-Wres},
are given in Table 2. Due to the
large energy scale there, the anomalous couplings are replaced by form factors
as  \eg $\lambda_Z \to \lambda_Z/(1+\hat s/\Lambda^2)$,
and the presented fits correspond to $\Lambda=1$ and 1.5 TeV.
\begin{table}[htbp]
\begin{center}
\begin{tabular}{||c|c|c||} \hline \hline
Condition          & $\Lambda$ = 1 TeV   & $\Lambda$ = 1.5 TeV   \\ \hline
$\Delta g^Z_1=\Delta \kappa_Z=0$
                   & $-0.53 < \lambda_Z < 0.56$
                                         & $-0.48 < \lambda_Z < 0.48$ \\
$\lambda_Z=\Delta \kappa_Z=0$
                   & $-0.57 < \Delta g^Z_1 < 0.76$
                                         & $-0.49 < \Delta g^Z_1 < 0.66$ \\
$\lambda_Z = 0$
                   & $-0.49< \Delta g_1^Z  = \Delta \kappa_Z < 0.66$
                                         & $-0.43 < \Delta g_1^Z
                                           = \Delta \kappa_Z < 0.57$  \\
$\lambda_Z=\Delta g^Z_1 =0 $
                   & $-2.0 < \Delta \kappa_Z < 2.4$
                                         & $-$                         \\
 \hline \hline
\end{tabular}
\caption{One-parameter  95\% C.L. fits from D0 \cite{D0-Wres}.}
\end{center}
\label{TEV-W-TGCtab}
\end{table}

\vspace{0.3cm}
As usual,   the $W^\pm$ TGC constraints  become  stronger  with energy.
Thus, even  stronger constrains are expected at LHC and ILC.
One additional reason for this, applying to  the specific operators
$\Op_W, \Op_{W\phi}, \tilde \Op_W$ in (\ref{W-TGC-op}), is that
they also produce  quartic couplings of the form  $WW\gamma\gamma$,
$WWZ\gamma$, $WWZZ$,  which may also be measured  \cite{Boudjema}.

Eventually, these constraints will become so strong, particularly for ILC, that
1-loop or higher SM results will be needed for correctly taking into account
the "SM-background".\\

\subsection{The on-shell anomalous triple neutral gauge couplings}

\begin{figure}[htb]
\[
\hspace{-0.5cm}\epsfig{file=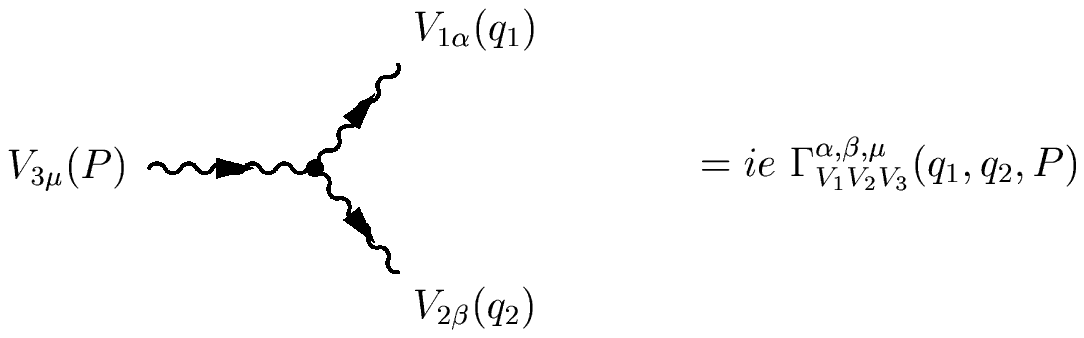,height=3.2cm, width=10.cm}
\]
\vspace*{-0.5cm}
\caption[1]{The  definition of the general triple neutral
gauge boson vertex, with $V_1,V_2$ taken on-shell, while $V_3$ is generally
off shell. }
\label{neutral-TGCfig}
\end{figure}

Using Fig.\ref{neutral-TGCfig} and \cite{Hagiwara, TGC-neutral},
the general triple neutral gauge vertex is written as
\bqa
\Gamma^{\alpha \beta \mu}_{ZZ V} (q_1, q_2, P)
&=& \frac{i (s-m_V^2)}{\mzd}
\left [ f_4^V (P^\alpha g^{\mu \beta}+P^\beta g^{\mu \alpha})
-f_5^V \epsilon^{\mu \alpha \beta \rho}(q_1-q_2)_\rho \right ]
~, \label{fZZ} \\
\Gamma^{\alpha \beta \mu}_{Z\gamma V} (q_1, q_2, P)
&=& \frac{ i (s-m_V^2)}{\mzd}
\Bigg \{ h_1^V (q_2^\mu g^{\alpha \beta}-q_2^\alpha g^{\mu \beta} )
+ \frac{h_2^V}{\mzd} P^\alpha [ (Pq_2) g^{\mu \beta}- q_2^\mu P^\beta ]
\nonumber \\
&-& h_3^V \epsilon^{\mu \alpha \beta \rho} q_{2\rho}
~-~\frac{h_4^V}{\mzd} P^\alpha \epsilon^{\mu \beta \rho
\sigma}P_\rho q_{2\sigma} \Bigg \}~ , \label{hZgamma}
\eqa
where $(V_3=Z,\gamma )$ is generally off-shell, while
the other two neutral gauge bosons are always on-shell.
If the NP scale is very high, all couplings in (\ref{fZZ},\ref{hZgamma})
are real. Singularities develop only if the NP scale is nearby.
The couplings $(f_5^V, h_3^V, h_4^V)$  respect CP,
while $(f_4^V, h_1^V, h_2^V)$ violate it. Finally,
the $(h_2^V, h_4^V)$-interactions
 may be relatively suppressed, since they are of higher dimension.

Based on \cite{LEPres},   the
fitted LEP ranges for the  $ZZ$-production couplings are indicated in
Table 3,  for  the cases that only
one  or  only two anomalous couplings are   possibly
non-vanishing.
The corresponding results for $Z\gamma$ production at LEP are given in
Table 4 \cite{LEPres}; while the D0 results
appear in Table\footnote{As in Table 2,
the anomalous couplings are replaced  in \cite{D0-neutral} by form factors
as   $h_i^V \to h_{i0}^V/(1+\hat s/\Lambda^2)^n$ with $n=3$ for $(i=1,3)$
and with $n=4$ for $(i=2,4)$.} 5.

\begin{table}[htbp]
\begin{center}
\renewcommand{\arraystretch}{1.3}
\begin{tabular}{|l||c|}
\hline
Parameter     & 95\% C.L.     \\
\hline
\hline
$f_4^\gamma$  & [$-0.17,~~+0.19$]  \\
\hline
$f_4^Z$       & [$-0.30,~~+0.30$]  \\
\hline
$f_5^\gamma$  & [$-0.32,~~+0.36$]  \\
\hline
$f_5^Z$       & [$-0.34,~~+0.38$]  \\
\hline
\end{tabular}
\begin{tabular}{|l||c|rr|}
\hline
Parameter     & 95\% C.L. & \multicolumn{2}{|c|}{Correlations} \\
\hline
\hline
$f_4^\gamma$  &[$-0.17,~~+0.19$] & $ 1.00$ & $ 0.07$\\
$f_4^Z$       &[$-0.30,~~+0.29$] & $ 0.07$ & $ 1.00$\\
\hline
$f_5^\gamma$  &[$-0.34,~~+0.38$] & $ 1.00$ & $-0.17$\\
$f_5^Z$       &[$-0.38,~~+0.36$] & $-0.17$ & $ 1.00$\\
\hline
\end{tabular}
 \label{ZZ-TGC-LEPtab}
 \caption[]{ The fitted parameters for the anomalous neutral TGC
from  the LEP $ZZ$ production \cite{LEPres}.
  Only the listed parameters are varied in each case;
  one in the left panel and two in the right one.
  In each case, the non-listed parameters  are  vanishing.}
\end{center}
\end{table}
\begin{table}[htbp]
\begin{center}
\renewcommand{\arraystretch}{1.3}
\begin{tabular}{|l||c|}
\hline
Parameter     & 95\% C.L.      \\
\hline
\hline
$h_1^\gamma$  & [$-0.056,~~+0.055$]  \\
\hline
$h_2^\gamma$  & [$-0.045,~~+0.025$]  \\
\hline
$h_3^\gamma$  & [$-0.049,~~-0.008$]  \\
\hline
$h_4^\gamma$  & [$-0.002,~~+0.034$]  \\
\hline
$h_1^Z$       & [$-0.13,~~+0.13$]  \\
\hline
$h_2^Z$       & [$-0.078,~~+0.071$]  \\
\hline
$h_3^Z$       & [$-0.20,~~+0.07$]  \\
\hline
$h_4^Z$       & [$-0.05,~~+0.12$]  \\
\hline
\end{tabular}
\begin{tabular}{|l||c|rr|}
\hline
Parameter  & 95\% C.L. & \multicolumn{2}{|c|}{Correlations} \\
\hline
\hline
$h_1^\gamma$  & [$-0.16,~~+0.05$]    & $ 1.00$ & $+0.79$ \\
$h_2^\gamma$  & [$-0.11,~~+0.02$]    & $+0.79$ & $ 1.00$ \\
\hline
$h_3^\gamma$  & [$-0.08,~~+0.14$]    & $ 1.00$ & $+0.97$ \\
$h_4^\gamma$  & [$-0.04,~~+0.11$]    & $+0.97$ & $ 1.00$ \\
\hline
$h_1^Z$       & [$-0.35,~~+0.28$]    & $ 1.00$ & $+0.77$ \\
$h_2^Z$       & [$-0.21,~~+0.17$]    & $+0.77$ & $ 1.00$ \\
\hline
$h_3^Z$       & [$-0.37,~~+0.29$]    & $ 1.00$ & $+0.76$ \\
$h_4^Z$       & [$-0.19,~~+0.21$]    & $+0.76$ & $ 1.00$ \\
\hline
\end{tabular}
 \label{Zgamma-TGC-LEPtab}
\caption[]{The fitted parameters for the anomalous neutral TGC
 from the LEP $Z\gamma$ production \cite{LEPres}.
  Only the listed parameters are varied in each case;
  one in the left panel and two in the right one.
  In each case, the non-listed parameters  are  vanishing.  }
\end{center}
\end{table}
\begin{table}[htbp]
\begin{center}
\begin{tabular}{||c|c|c||} \hline \hline
Coupling               & $\Lambda =$~750~GeV & $\Lambda =$~1~TeV\\ \hline
$|\Re {\it e} (h_{10, 30}^{Z})|$, $|\Im {\it m} (h_{10, 30}^{Z})|$          & 0.24                & 0.23 \\
$|\Re {\it e} (h_{20, 40}^{Z})|$, $|\Im {\it m}(h_{20, 40}^{Z})|$           & 0.027               & 0.020\\
$|\Re {\it e} (h_{10, 30}^{\gamma})|$, $|\Im {\it m}(h_{10, 30}^{\gamma})|$ & 0.29                & 0.23 \\
$|\Re {\it e} (h_{20, 40}^{\gamma})|$, $|\Im {\it m}(h_{20, 40}^{\gamma})|$ & 0.030               & 0.019\\
\hline \hline
\end{tabular}
\label{Zgamma-TGC-D0tab}
\caption{The fitted parameters for the anomalous neutral TGC
from  the D0 $Z\gamma$ production \cite{D0-neutral}.
  Only the listed parameters are varied in each case,
  which are taken to be either purely real or purely imaginary.
  In each case, the non-listed parameters  are  vanishing. }
\end{center}
\end{table}

\vspace{1cm}
The  overall conclusion on the basis of Fig.\ref{LEP-W-TGCfig} and
Tables 1-5, is that
\underline{no indication} for any anomalous TGC exists at present.\\

\section{Helicity Conservation and its possible violation. }

We next turn to the Helicity Conservation (HC) property,
restricting to processes of even order in the Yukawa
couplings \cite{helicity}. Simple rules are then obtained,
that generically test the presence of  anomalous couplings
for any two-body process at high energies and fixed angles \cite{helicity}.
Thus, denoting its  helicity  amplitudes
by $F(a_{\lambda_1}b_{\lambda_2}\to c_{\lambda_3}d_{\lambda_4})$,
the  allowed  helicities at  asymptotic $(s,|t|,|u|)$-values are constrained
as
\bq
\lambda_1+\lambda_2=\lambda_3+\lambda_4  ~~, \label{HC1-con}
\eq
unless the two initial (or  final) particles are fermions
and  the other two  bosons,  where the stronger relation
\bq
\lambda_1+\lambda_2=\lambda_3+\lambda_4=0  ~~ \label{HC2-con}
\eq
is imposed.

Particularly for   transverse gauge bosons,
the  structure for the  asymptotically non-vanishing two-body
helicity amplitudes implied by HC  is
\bqa
&& F(f_{\lambda_f}f'_{\lambda_{-f}}\to V_{\lambda_V}V'_{\lambda_{-V}})~~,~~
F( V_{\lambda_V}V'_{\lambda_{-V}}\to f_{\lambda_f}f'_{\lambda_{-f}})~~,
\label{HC3-con} \\
&&  F( V_{\lambda_V}V'_{\lambda_{-V}}\to  \phi  \phi')~~,~~
F( \phi  \phi' \to V_{\lambda_V}V'_{\lambda_{-V}} )~~, \label{HC4-con} \\
&& F(V_{\lambda_V} f_{\lambda_f} \to V'_{\lambda_{V}}f'_{\lambda_{f}})~~,~~
 F(V_{\lambda_V} \phi  \to V'_{\lambda_{V}}\phi')~~,~~ \label{HC5-con}
\eqa
where  by $f,\phi ,V$ we denote fermion, scalar or vector particles
respectively.

Equations (\ref{HC1-con}, \ref{HC2-con}) remain
of course true even in the presence of longitudinal vector
bosons\footnote{Obviously, the helicities
of a fermion are $\pm 1/2$,
of a vector boson $(\pm 1,~ 0)$, while  they are vanishing
for a  scalar particle. The longitudinal vector boson helicity
is also denoted below by $L$.}. For the  vector boson amplitudes denoted as
$ F(V^1_{\lambda_1}V^2_{\lambda_2}\to V^3_{\lambda_3} V^4_{\lambda_4})
\equiv F_{\lambda_1\lambda_2\lambda_3\lambda_4}$, they also imply relations like
\bqa
&&F_{+++-}=F_{++-+}=F_{+-++}=F_{-+++}=F_{---+}\nonumber\\
&&=F_{++LL}=F_{-+-L}=F_{+--L}=F_{++L-} \simeq 0~~, \label{HC-gauge}
\eqa
since all HC-violating amplitudes
should necessarily vanish  at high $(s,|t|,|u|)$.

The most important ingredient for the validity of HC in either SM or MSSM,
is \underline{renormalizability} \cite{helicity}.

For processes involving fermions or scalars only, HC holds
at a diagram-by-diagram basis.
For  gauge involving amplitudes though, the situation is more subtle.
Large  cancellations among the various diagrams are needed in order to achieve HC.
This way, HC is established  at the Born level in both SM and MSSM.
When going beyond this though, intriguing differences between SM and  MSSM
appear, which we summarize below.

In  SM, HC is only valid  up to the $\ln^2$ and $\ln$ terms of
the 1-loop corrections, provided $(s,|t|,|u|)\gg (\mwd, m^2_H)$.
The theorem is easier to be established for processes driven
by a non-vanishing Born contribution. In any case, it has been checked
explicitly to the leading log accuracy, for
$(e^-e^+ \to \gamma \gamma, ~ ZZ, ~\gamma Z, ~W^-W^+)$ using \cite{eeVV}, and
$ (\gamma \gamma \to ZZ, ~ \gamma Z, ~ ZZ)$ using \cite{Laz-gg, Laz-gZ, Laz-ZZ}.
Constant high energy
contributions  in SM though, usually violate HC.

In MSSM, HC is valid  to \underline{all} orders in the gauge
and Yukawa couplings, for any two-body process,
 at $(s,|t|,|u|)\gg M_{\rm SUSY}^2$  \cite{helicity}.
Constant contributions  respect it also!

SUSY somehow knows of the cancellations among the various
diagrams describing the gauge  boson involving processes.
The reason for this is that,
at high energies SUSY associates each gauge boson of a definite helicity,
to a corresponding gaugino carrying a helicity of the same sign.
Since,   HC is valid for fermions at a
diagram-by-diagram basis; it should  be valid for gauge bosons also.
In the general proof, masses have been neglected \cite{helicity}.

The validity of HC, even  for the constant asymptotic
contributions in MSSM, has also been observed  in
$ \gamma \gamma \to ZZ, ~ \gamma Z, ~ ZZ$, for which the exact 1-loop
results are known \cite{helicity, Laz-gZ, Laz-ZZ}.

\begin{table}[htbp]
\begin{center}
\begin{small}
\begin{tabular}{||c c c c c||} \hline \hline
    & $\tau=\tau'=\pm 1$ &  $\tau=-\tau'=\pm 1$ &
    $\tau=\tau'=0$  & $\tau=0,\tau'=\pm 1, \epsilon=1$  \\
    & $-\frac{\ed}{2}\lambda \sintheta $ &  $-\frac{\ed}{2}\lambda \sintheta $ &
     $-\frac{\ed}{2}\lambda \sintheta $  & $-\ed\lambda
     \frac{(\tau'\costheta -2\lambda ) }{2\rd}$                  \\   \hline
$\delta_Z$ & $-2\delta_Z (a-2 b\lambda)$ & 0&
$-\frac{s}{\mwd} \delta_Z (a-2 b\lambda)$ &
$-\frac{\sqrt{s}}{\mw} 2\delta_Z (a-2 b\lambda) $ \\
$ x_\gamma~,x_Z$ &0 & 0 &
$\frac{s}{\mwd}[x_\gamma-x_Z(a-2 b\lambda)] $ &
$\frac{\sqrt{s}}{\mw}[x_\gamma-x_Z(a-2 b\lambda)] $ \\
$y_\gamma~,y_Z$ & $\frac{s}{\mwd}[y_\gamma-y_Z(a-2 b\lambda)] $ & 0 & 0 &
$\frac{\sqrt{s}}{\mw}[y_\gamma-y_Z(a-2 b\lambda)]$ \\
$z_Z$ & 0 & 0 & 0 &
$-\Big (\frac{\sqrt{s}}{\mw}\Big )^3 z_Z(a-2 b\lambda)\tau'$ \\
$z'_1$ & 0 & 0 & 0 &
$-i \frac{\sqrt{s}}{\mw} z'_1(a-2 b\lambda)\epsilon $ \\
$z'_2$ & $i z'_2 2 \tau (a-2 b\lambda) $ & 0 & 0 &
$i \frac{\sqrt{s}}{\mw} z'_2\tau'(a-2 b\lambda)\epsilon $ \\
$z'_3$ & $-i z'_3 2 \tau (a-2 b\lambda)\frac{s}{\mwd} $ & 0 & 0 & 0 \\
\hline \hline
\end{tabular}
\caption{The leading  large-s anomalous contribution
to $F(e^-_\lambda e^+_{-\lambda} \to W^-_\tau W^+_{\tau'})$ \cite{Bilenky}.
The helicity amplitudes are obtained from each
column by multiplying the factor on top, with the
 sum of  all its  elements. The first column indicates the
 anomalous couplings contributing. The amplitudes for $\tau=\pm 1, \tau'=0$
are obtained from the last column by substituting there $\tau' \to -\tau $ and
$\epsilon=-1$. }
\end{small}
\label{eeWW-TGC-tab}
\end{center}
\end{table}

We next turn to the anomalous contributions to the asymptotic two-body
amplitudes. Since the most we can expect about such couplings is
that they are very small, we always calculated their contribution at the
Born level. For $F(e^-_\lambda e^+_{-\lambda} \to W^-_\tau W^+_{\tau'})$,
the  complete asymptotic anomalous contributions to the helicity amplitudes
are given in Table 6 \cite{Bilenky}, where (\ref{W-TGC}) and
the definitions
\bqa
&& a=\frac{-1+4\swd}{4\sw\cw} ~~,~~ b=\frac{-1}{4\sw\cw} ~~,~~
 \delta_Z = \frac{\cw}{\sw} \delta g_1^Z  ~~, \nonumber \\
&& x_\gamma = \delta\kappa_\gamma ~~,~~
x_Z=(\delta \kappa_Z-\delta g_1^Z)\frac{\cw}{\sw}~~, ~~
y_\gamma=\lambda_\gamma ~~,~~ y_Z=\lambda_Z \frac{\cw}{\sw}~~
\eqa
are used.
The  CP violating couplings $(z'_1,z'_2,z'_3)$ in the last three rows
of Table 6, are linear combinations of the couplings
$(g_4^Z,\tilde \kappa_Z, \tilde \lambda_Z)$ defined in  (\ref{W-TGC})
 \cite{Bilenky}.

As seen from Table 6, none of the TGC in (\ref{W-TGC}),
 respects HC. Thus, bounds on the ratios
 \bqa
\frac{|F(e^-_\lambda e^+_{-\lambda} \to W^-_0 W^+_{\pm 1})|}
{|F(e^-_\lambda e^+_{-\lambda} \to W^-_{\pm 1} W^+_{\mp 1})|} &,&
\frac{|F(e^-_\lambda e^+_{-\lambda} \to W^-_{\pm 1} W^+_0)|}
{|F(e^-_\lambda e^+_{-\lambda} \to W^-_{\pm 1} W^+_{\mp 1})|} ~~, \nonumber \\
\frac{|F(e^-_\lambda e^+_{-\lambda} \to W^-_{\pm 1} W^+_{\pm 1})|}
{|F(e^-_\lambda e^+_{-\lambda} \to W^-_{\pm 1} W^+_{\mp 1})|} &,&
\frac{|F(e^-_\lambda e^+_{-\lambda} \to W^-_{0} W^+_{0})|}
{|F(e^-_\lambda e^+_{-\lambda} \to W^-_{\pm 1} W^+_{\mp 1})|} ~~, \nonumber
\eqa
measured at the high energy part of Linear Collider (ILC),
could constrain all anomalous couplings.\\

As  further   examples of  anomalous HC violations
in  other  2-body processes, we give in
(\ref{eeWW}-\ref{gWZW}), the SM and $\Op_W$
contributions to the high energy helicity amplitudes\footnote{In
(\ref{eeWW}-\ref{gWZW}), $s$ denotes the subprocess c.m. squared energy.}
\cite{HCviol}; compare (\ref{W-TGC-d6}, \ref{W-TGC-op}).
In all cases, the HC violating amplitudes,
indicated through a double arrow in the left hand sides of
(\ref{eeWW}-\ref{gWZW}), are determined by the
anomalous interactions. These are\\

\hspace{2cm} \fbox{ $ d\bar d,\ u\bar u\to W^- W^+$}
\vspace*{-0.3cm}
\bqa
&&  \Rightarrow ~~ F^L_{++} = F^L_{--}=\:\tt\: {\ed\over4\swd}\lsm\sintheta
\nonumber\\
&& F^L_{+-} = -{\ed\over2\swd}\sintheta\:{(1-\tt\costheta)\over1+\costheta}\ \ \ ;\ \
F^L_{-+} = {\ed\over2\swd}\sintheta\:{(1-\tt\costheta)\over1-\costheta} \nonumber \\
&& F^L_{LL}= \tt{\ed\over2\cwd}\sintheta\:(|Q|-1+{1\over2\swd})\nonumber \\
&& F^R_{LL}= Q{\ed\over2\cwd}\sintheta ~~ ,\label{eeWW}
\eqa
where $Q$ is the quark charge, \\

\hspace{3.cm} \fbox{ {\bf $d\bar u\to  W^- Z$ }}
\vspace*{-0.3cm}
\bqa
&& \Rightarrow ~~  F^L_{++} = F^L_{--}=-\:{\ed \over2\rd}\: {c_W\over s^2_W}\:
\lsm\sintheta\nonumber\\
&& F^L_{+-} = -\:{\ed\over\rd
c_W\swd}\:\:{\sintheta\over1+\costheta}\ \bigm(\cwd\costheta-{\swd\over3}\bigm)
\nonumber\\
&& F^L_{-+} ={\ed\over\rd c_W\swd}\:\: {\sintheta\over1-\costheta}\
\bigm(\cwd\costheta-{\swd\over3}\bigm)\nonumber\\
&& F^L_{LL}= -\:{\ed\over2\rd\, \swd}\:\sintheta  ~~, \label{dbuWZ}
\eqa\\

\hspace{3.cm} \fbox{ {\bf $d\bar u\to  W^- \gamma$ }}
\vspace*{-0.3cm}
\bqa
&& \Rightarrow ~~  F^L_{++} = F^L_{--}=\:-\:{\ed \over2\rd\, s_W}\:\sintheta\lsm\nonumber \\
&& F^L_{+-} = -\:{\ed\over\rd\, s_W}\ \
 {\sintheta\over1+\costheta}\ \bigm(\costheta+{1\over3}\bigm)\nonumber\\
&& F^L_{-+} = {\ed\over\rd\, s_W}\ \
 {\sintheta\over1-\costheta}\ \bigm(\costheta+{1\over3}\bigm) ~~, \label{dbuWg}
 \eqa \\

\hspace{1cm}\fbox{ {\bf $ \gamma\gamma\to W W$ }}
\vspace*{-0.3cm}
\bqa
&& F_{++++}=F_{----}=\ {8\ed\over\sin^2\theta}\nonumber\\
&& \Rightarrow ~~  F_{+++-}=F_{++-+}=\ F_{+-++}=\ F_{-+++}=  \nonumber \\
&& \Rightarrow ~~  F_{---+}=F_{--+-}=\ F_{-+--}=\ F_{+---}=\ -\ed\,\lsm\nonumber \\
&& F_{+--+}=F_{-++-}=\ed(1-\costheta)\biggm\{{2\over1+\costheta}+{3+\costheta\over16}
\lsm^2\biggm\}\nonumber \\
&& \Rightarrow ~~  F_{++--}=F_{--++}=\ed\lsm
  \biggm\{-2\ +\ {3-\cos^2\theta\over8}\lsm\biggm\}\nonumber \\
&& F_{+-+-}=F_{-+-+}=\ -\ed\ (1+\costheta)
\biggm\{{2\over\costheta-1}+{(\costheta-3)\over16}\ \lsm^2\biggm\}\nonumber \\
&& F_{+-LL}=F_{-+LL}=2\ed ~~, \label{ggWW}
\eqa\\

\hspace{1cm}\fbox{ {\bf $ \gamma W\to \gamma W$ }}
\vspace*{-0.3cm}
\bqa
 && F_{++++}=F_{----}=\:-\:\ed\biggm\{ {4\over1+\costheta}\:+\:\lsm^2\
{\costheta\over4}\biggm\} \nonumber\\
&& \Rightarrow ~~ F_{+++-}=F_{++-+}=\ F_{+-++}=\ F_{-+++}=  \nonumber\\
&& \Rightarrow ~~ F_{---+}=F_{--+-}=\ F_{-+--}=\ F_{+---}= \ed\ {(1-\costheta)\over2}\ \lsm
\nonumber\\
&& F_{+--+}=F_{-++-}=\:-\:\ed\ {(1-\costheta)^2\over1+\costheta}\nonumber\\
&& F_{+-+-}=F_{-+-+}=\:-\:\ed\ (1+\costheta)\biggm\{1+{3-\costheta\over16}\
\lsm^2\biggm\}\nonumber\\
&&  \Rightarrow ~~ F_{++--}=F_{--++}=\ed\ \lsm
\biggm\{1-\costheta-{(3+6\costheta-\cos^2\theta)\over16}\ \lsm\biggm\}\nonumber \\
&& F_{+L+L}=F_{-L-L}=-2\ed ~~. \label{gWgW}
\eqa\\

\hspace{-1cm} \fbox{ {\bf $ \gamma W\to Z W$ }}\\[0.5cm]
The purely transverse amplitudes are identical to  those
for $\gamma W\to \gamma W$ in (\ref{gWgW}),
provided we replace $\ed\to \ed c_W/s_W $. The
amplitudes involving longitudinal bosons are
 \bqa
&& \Rightarrow ~~ F_{++LL}=F_{--LL}=\ {\ed\over4s_W}\costheta\ \lsm\nonumber\\
&& F_{+-LL}=F_{-+LL}=\ -\ {\ed\over2s_W}\ (1-\costheta) \nonumber\\
&& \Rightarrow ~~ F_{+LL-}=F_{-LL+}=\ {\ed\over8s_W}\ (\costheta-3)\ \lsm\nonumber\\
&& F_{+LL+}=F_{-LL-}=\ -\ {\ed\over s_W}\ {(\costheta-1)\over\costheta+1}
~~ . \label{gWZW}
\eqa

Eqs.(\ref{eeWW}-\ref{gWZW}) also indicate that  the $\Op_W$ contributions
to the helicity conserving amplitudes are  always quadratic in $\alpha_W$,
and therefore suppressed. Thus, measurements of HC violations should
be very sensitive to $\Op_W$. Similar results apply
also to  any other anomalous interaction.\\

\section{Conclusions}

There is no real indication at present that any anomalous couplings exist.
This is supported also by the LEP \cite{LEPres} and
Tevatron \cite{D0-Wres, D0-neutral}
results already available.
At the high energies accessible to LHC and ILC, we would expect
these constraints to  become stronger.

Since the energies available at LHC and ILC are  very high,
the subprocess conditions  $( s,|t|,|u|)\gg (\mwd~, m_H^2)$ should be satisfiable,
so that HC is respected by the electroweak sector
of SM to a high accuracy\footnote{At least
if no top contributions are important.}.
In any case, we would  expect HC to be respected
 to the  1-loop leading $(\ln^2,\ln)$ terms  in
\bqa
 &&  q\bar q \to gg~,~ g\gamma ~,~ gZ~,~ gW~,~ \gamma\gamma ~,~
\gamma Z~,~ ZZ~,~ W^+W^- ~,~ \gamma W~,~  ZW~~, \nonumber  \\
&&   gq \to gq ~,~ \gamma q ~,~ Zq ~,~ Wq ~, \nonumber \\
&&  gg \to gg ~,~ q\bar q ~, \nonumber \\
&&  e^+e^-\to \gamma\gamma ~,~ \gamma Z~,~ZZ~,W^+W^-~ ,\nonumber \\
 && \gamma e\to \gamma e ~,~ Ze~,~ W\nu ~,~ \nonumber \\
 && \gamma \gamma \to  f\bar f ~,~ \gamma\gamma ~,~
\gamma Z~,~ ZZ~,~ W^+W^-~.~ \label{HC-SM}
\eqa

If SUSY is realized in Nature and
$(s,|t|,|u|)\gg M_{\rm SUSY}^2$ is also satisfied within the LHC or ILC range,
then HC should be valid  for all processes
in  (\ref{HC-SM}), as well as  in
\bqa
&& gg\to \tilde g\tilde g~,~ \tilde q\bar {\tilde q}~~, \nonumber \\
&& e^-e^+ \to \tilde f \bar {\tilde  f}~,~ \tchi^+ \tchi^- ~,
H^+H^-~,~ H^0H^{\prime 0}~~, \nonumber \\
&& \gamma\gamma \to \tilde f \bar {\tilde  f}~,~ \tchi^+ \tchi^- ~,
H^+H^-~,~ H^0H^{\prime 0} ~~, \label{HC-MSSM}
\eqa
where $H^0$ denotes any of the neutral Higgs particles in MSSM.

In either case,  detail studies  may identify those of the above processes,
 which are
the most suitable for excluding the anomalous contributions  violating HC.
Thus, searching for HC violations may be
 a useful way for constraining the
anomalous couplings, and at the same time,  any effectively non-renormalizable way
of going beyond the standard model. Some realizations of extra large dimensions
may  fall in this later category.

\end{document}